\documentclass[a4paper,twocolumn,11pt,accepted=2020-10-16]{quantumarticle}
\pdfoutput=1
\usepackage[utf8]{inputenc}
\usepackage[english]{babel}
\usepackage[T1]{fontenc}
\usepackage{amsmath}
\usepackage{amssymb}
\usepackage[numbers,sort&compress]{natbib}

\usepackage{dsfont}
\usepackage{graphicx}
\usepackage{xspace}
\usepackage{xcolor}
\usepackage{acronym}
\usepackage{hyperref}

\usepackage{placeins}

\newacro{GKSL}[GKSL]{Gorini--Kossakowski--Sudarshan--Lindblad}
\newacro{RWA}[RWA]{rotating wave approximation}
\newacro{HOPS}[HOPS]{hierarchy of pure states}
\newacro{SD}[SD]{spectral density}
\newacro{QOME}[QOME]{quantum-optical master equation}
\newacro{CGME}[CGME]{coarse-graining master equation}
\newacro{RFE}[RFE]{Redfield equation}
\newacro{GAME}[GAME]{geometric-arithmetic master equation}
\newacro{PRWA}[PRWA]{partial rotating wave approximation}
\newacro{BCF}[BCF]{bath correlation function}

\newcommand{\imag}{\mathrm{i}}

\newcommand{\sys}{\mathrm{sys}}

\newcommand{\env}{\mathrm{env}}

\newcommand{\hc}{\mathrm{h.c.}}
\newcommand{\bra}[1]{\langle #1 |}
\newcommand{\ket}[1]{|{#1} \rangle}

\newcommand{\il}[3]{\int_{#1}^{#2}\mathrm{d}#3\,}
\newcommand{\w}{\omega}

\begin{document}
\title{Environmentally Induced Entanglement -- Anomalous Behavior in the Adiabatic Regime}

\author{Richard Hartmann}
\email{richard.hartmann@tu-dresden.de}
\affiliation{Institut für Theoretische Physik, Technische Universität Dresden, D-01062 Dresden, Germany}
\orcid{0000-0002-8967-6183}

\author{Walter T. Strunz}
\affiliation{Institut für Theoretische Physik, Technische Universität Dresden, D-01062 Dresden, Germany}
\orcid{0000-0002-7806-3525}

\begin{abstract}

Considering two non-interacting qubits in the context of open quantum systems, it is well known that their common environment may act as an entangling agent.
In a perturbative regime the influence of the environment on the system dynamics can effectively be described by a unitary and a dissipative contribution.
For the two-spin Boson model with (sub-) Ohmic spectral density considered here, the particular unitary contribution (Lamb shift) easily explains the buildup of entanglement between the two qubits.
Furthermore it has been argued that in the adiabatic limit, adding the so-called counterterm to the microscopic model compensates the unitary influence of the environment and, thus, inhibits the generation of entanglement.
Investigating this assertion is one of the main objectives of the work presented here.
Using the hierarchy of pure states (HOPS) method to numerically calculate the exact reduced dynamics, we find and explain that the degree of inhibition crucially depends on the parameter $s$ determining the low frequency power law behavior of the spectral density $J(\omega) \sim \omega^s e^{-\omega/\omega_c}$.
Remarkably, we find that for resonant qubits, even in the adiabatic regime (arbitrarily large $\omega_c$), the entanglement dynamics is still influenced by an environmentally induced Hamiltonian interaction. Further, we study the model in detail and present the exact entanglement dynamics for a wide range of coupling strengths, distinguish between resonant and detuned qubits, as well as Ohmic and deep sub-Ohmic environments.
Notably, we find that in all cases the asymptotic entanglement does not vanish and conjecture a linear relation between the coupling strength and the asymptotic entanglement measured by means of concurrence.
Further we discuss the suitability of various perturbative master equations for obtaining approximate entanglement dynamics.

\end{abstract}

\maketitle

\section{Introduction}

Entanglement -- this very peculiar kind of correlation, not occurring in the classical world, is known to be a fragile property with respect to environmental influences \cite{ZurekDecoherenceeinselectionquantum2003, DurStabilityMacroscopicEntanglement2004, AlmeidaEnvironmentInducedSuddenDeath2007, YuSuddenDeathEntanglement2009}.
On the other hand, such an environmental interaction features also the capability to induce entanglement \cite{BraunCreationEntanglementInteraction2002, KimEntanglementinducedsinglemode2002, IsarEntanglementGenerationEvolution2009, MazzolaSuddenDeathSudden2009}, which can even last for arbitrary long times \cite{BenattiEntanglingoscillatorsenvironment2006, ZellDistanceDependenceEntanglement2009, SahrapourTunnelingDecoherenceEntanglement2013}.
The investigation of these competing environmental effects \cite{DubeDynamicsPairInteracting1998, ZyczkowskiDynamicsquantumentanglement2001, BasharovDecoherenceEntanglementRadiative2002, JakobczykEntanglingtwoqubits2002, SchneiderEntanglementsteadystate2002, LendiDaviestheoryreservoirinduced2006, PazDynamicsEntanglementTwo2008, SahrapourTunnelingDecoherenceEntanglement2013, KastBipartiteentanglementdynamics2014, AolitaOpensystemDynamicsEntanglement2015, DengDynamicsTwospinSpinboson2016, EasthamBathinducedcoherencesecular2016} adds to the understanding of entanglement and decoherence in the context of open quantum systems which is not only of importance from a fundamental point of view but is also most relevant for fields like quantum computation, communication and metrology.

Albeit the vast amount of publications in this field, sophisticated results for the microscopic model of a system and its environment are rare, even for the simple case of two qubits coupled to a common (sub-) Ohmic environment.
Therefore we investigate the two-qubit entanglement by means of the numerically exact \ac{HOPS} method \cite{SuessHierarchyStochasticPure2014, ZhangNonPerturbativeCalculationTwoDimensional2016, HartmannExactOpenQuantum2017, ZhangFlexibleschemetruncate2018, HartmannExactopenquantum2019} with two major objectives in mind.
First, we draw conclusions about the general applicability of various perturbative approaches.
It is in line with previous results \cite{HartmannAccuracyAssessmentPerturbative2020} that the \ac{RFE} is highly favorable.
As expected, our investigation confirms that in the perturbative regime the entanglement generation is well modeled by the environmentally induced Lamb shift Hamiltonian mediating an effective qubit-qubit interaction.
Further, it has also been argued that this unitary environmental effect on the system may be canceled by the so-called counterterm in the adiabatic limit \cite{WeissQuantumDissipativeSystems2008}.
Therefore, we secondly investigate the influences of the counterterm on the entanglement dynamics and find that it sensitively depends on the parameter $s$ of the \ac{SD}.
Whereas in the Ohmic case the inhibition of entanglement is well observed, in the deep sub-Ohmic regime ($s \lesssim 0.3$) a significantly larger cutoff frequency $\omega_c$ is required to see the same effect.
Notably, even under the assumption that the environment acts dissipatively only entanglement may still be generated \cite{BenattiEnvironmentInducedEntanglement2003}, however, much smaller in magnitude and on a slower time scale.

The model under consideration extends the prominent spin-boson model \cite{LeggettDynamicsdissipativetwostate1987} to two qubits
\begin{equation}
\begin{gathered}
  H = H_{\sys} + L\otimes \sum_{\lambda} g_{\lambda} (a_{\lambda} + a_{\lambda}^{\dagger}) + \sum_{\lambda} \omega_{\lambda} a^{\dagger}_{\lambda} 
  a_{\lambda} \\    
  H_{\sys} = -\frac{\omega_{A}}{2} \sigma_{x}^{A} - \frac{\omega_{B}}{2} \sigma_{x}^{B} \qquad 
  L = \frac{1}{2} (\sigma_{z}^{A} + \sigma_{z}^{B}) \quad .
  \label{eqn:hamiltonian}
\end{gathered}
\end{equation}
Each qubit is modeled by the Pauli matrix $\sigma_x$ with tunneling frequency $\omega$.
They are coupled via $\sigma_z$ to a common bosonic environment where $a$ ($a^\dagger$) denotes the bosonic annihilation (creation) operator. 
Note, units are chosen such that $\hbar$ and $k_B$ become unity.
For the continuous environment we choose a (sub-) Ohmic \ac{SD}
\begin{equation}
  \pi \sum_\lambda g_\lambda^2 \delta(\omega - \omega_\lambda) = J(\omega) = \frac{\pi}{2}\alpha\omega_c^{1-s}\omega^s e^{-\omega/\omega_c} \quad ,
\end{equation}
where $\alpha$ denotes the dimensionless coupling strength, $\omega_c$ the cutoff frequency of the exponential cutoff and the power $s$ allows to distinguish between the sub-Ohmic ($s<1$), Ohmic ($s=1$) and super-Ohmic ($s>1$) regime.

For such a microscopic model time local master equations provide perturbative results in the weak coupling regime and/or for a fast decaying \ac{BCF}.
In the case of two resonant qubits ($\omega_A = \omega_B$), the widely used \ac{QOME} \cite{DaviesMarkovianmasterequations1974, BreuerTheoryOpenQuantum2007} of \ac{GKSL} form \cite{ChruscinskiBriefHistoryGKLS2017} reveals that the relaxation dynamics is accompanied by an environmentally induced Hermitian coupling between the two parties \cite{BreuerTheoryOpenQuantum2007, KryszewskiMasterequationtutorial2008, WhitneyStayingpositivegoing2008, McCutcheonLonglivedspinentanglement2009, HartmannAccuracyAssessmentPerturbative2020}. 
This effective interaction easily explains the entanglement generation, even in the absence of a direct coupling \cite{LendiDaviestheoryreservoirinduced2006, SolenovExchangeinteractionentanglement2007, McCutcheonLonglivedspinentanglement2009}.
However, for the more general case of two detuned qubits the \ac{QOME} does not show entanglement generation \cite{BenattiEntanglingtwounequal2010, MajenzCoarseGrainingCan2013}.
At first glance this seems contradictory since a small change in the system Hamiltonian should result in a small change in the dynamics, too.
But, since the \ac{QOME} relies on the \ac{RWA}, the slow phase caused by a small detuning requires a substantially weaker coupling for the \ac{RWA} to hold as compared to the resonant case (see \cite{HartmannAccuracyAssessmentPerturbative2020}).

We show by means of numerical results that if the coupling strength is not too small two detuned qubits become entangled, too.
We further investigate the question to what extent time local master equations, other than the \ac{QOME}, approximate the exact entanglement dynamics of two detuned qubits sufficiently well.

In addition, we find from the exact dynamics that even in the weak coupling regime the entanglement of the asymptotic state does not vanish irrespectively of the detuning.
The correct asymptotic value cannot be obtained by any of the time local master equations considered here.

To further enlighten the properties of environmentally induced entanglement we investigate the case where the environmentally mediated unitary qubit-qubit interaction is suppressed.
A similar study based on \ac{GKSL} type master equations has shown that if the bath induced unitary interaction is omitted, the action of the dissipator alone can result in entanglement generation as well \cite{BenattiEnvironmentInducedEntanglement2003}.

By means of the full microscopic model it is, a priori, not clear how to implement this scenario since the exact form of the induced unitary interaction is not known.
In the adiabatic limit, however, where the bath oscillators react instantaneously to the system dynamics \cite{WeissQuantumDissipativeSystems2008}, it takes the form
\begin{equation}
  - L^2 \sum_\lambda \frac{g_\lambda^2}{\omega_\lambda} = - \frac{\alpha \omega_c}{2} \Gamma(s) L^2 =: - H_c
\end{equation}
which defines the so-called ``counterterm'' ($\Gamma$ denotes the gamma function).
Based on this reasoning, adding the counterterm $H_c$ to the microscopic Hamiltonian [Eq. \eqref{eqn:hamiltonian}] is assumed to suppress the bath induced unitary interaction \cite{OrthDynamicssynchronizationquantum2010, KastBipartiteentanglementdynamics2014} and, thus, inhibit the buildup of entanglement.
However, the adiabatic assumption should be questioned for the deep sub-Ohmic regime where the low frequency modes are especially important.
We find and explain that for finite $\omega_c$ including the counterterm does not fully suppress the entanglement generation.
In addition we report the interesting phenomenon that for resonant qubits even in the limit $\omega_c \rightarrow \infty$ the truly adiabatic regime is never reached.
That is, a significant amount of entanglement will always be generated (and diminish afterwards) via an environmentally induced unitary interaction even in the presence of the renormalizing counter term.

The two-spin Boson model considered here with (sub-) Ohmic \ac{SD} and exponential cutoff has also been investigated by means of other numerical methods.
Unfortunately, the entanglement dynamics obtained using path integral Monte Carlo (PIMC) techniques \cite{KastBipartiteentanglementdynamics2014} does not agree with our numerical and analytical results.
As of missing information in Ref. \cite{SahrapourTunnelingDecoherenceEntanglement2013} we were not able to connect our calculations to results obtained from the quasi-adiabatic path integral (QUAPI) method.
However, the time-evolving matrix product operator (TEMPO) algorithm \cite{StrathearnEfficientNonMarkovianQuantum2018}, an advancement of QUAPI, gives results consistent with ours.

\section{Induced Entanglement Dynamics}

\begin{figure}[tb]
  \includegraphics[width = \columnwidth]{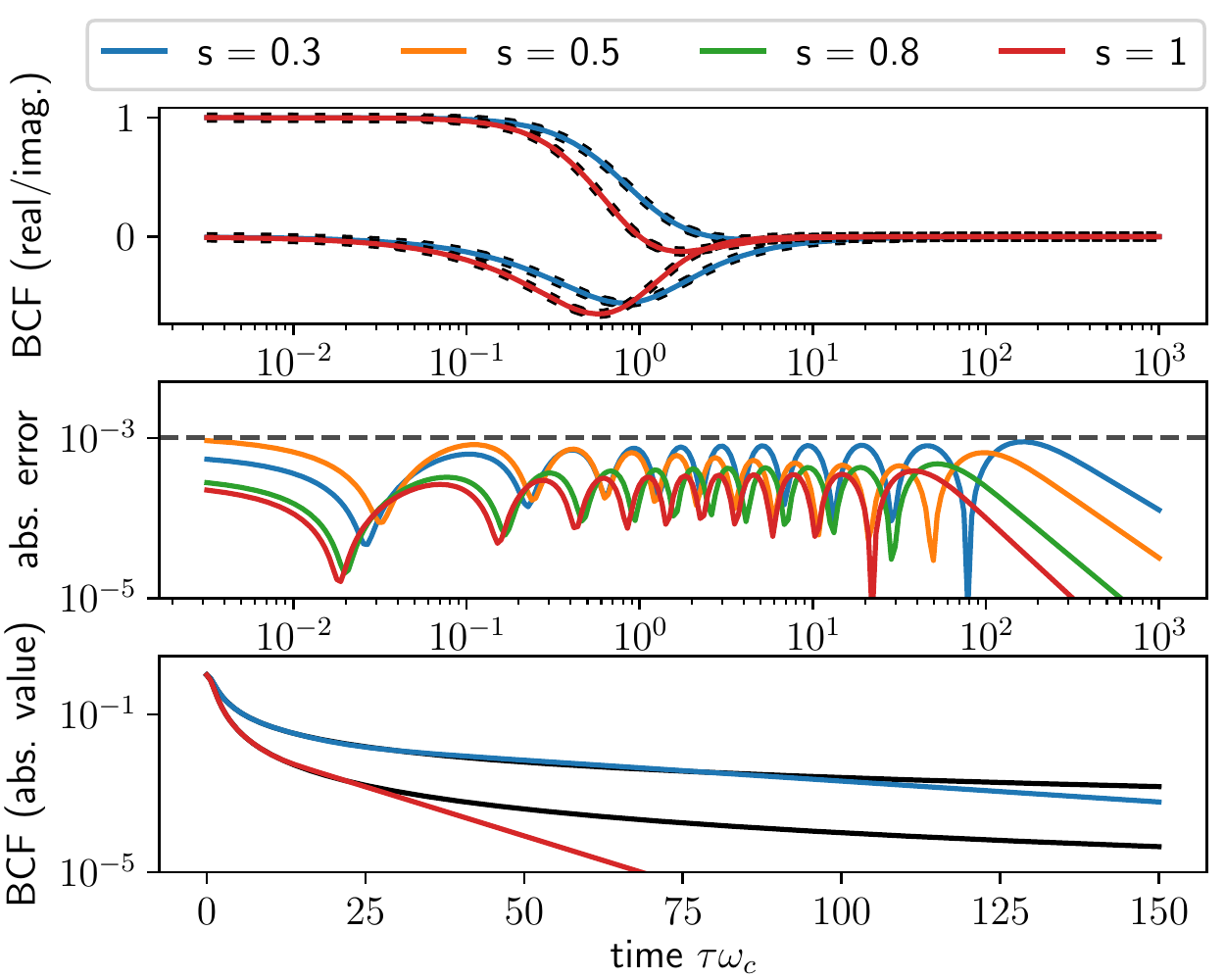}
  \caption{Details on the approximation of the (sub-) Ohmic \ac{BCF} in terms of a sum of exponentials are shown. Colored lines refer to the approximation and black lines to the exact \ac{BCF}. For the various parameters $s$ each approximation has a maximum absolute error of $10^{-3}$ (see the middle panel). Consequently, deviations are not visible when plotting the real/imaginary part on a linear scale (upper panel). However, showing the absolute value on a logarithmic scale reveals how the sum of exponentials mimics the algebraic decay up to a certain time (lower panel).}
  \label{fig:OhmFit}
\end{figure}

As a measure of the two-qubit entanglement we use concurrence \cite{HorodeckiQuantumentanglement2009}.
To examine its dynamics, the reduced dynamics of the two-spin boson model specified in Eq. \eqref{eqn:hamiltonian} is obtained numerically, solving the non-linear variant of the non-Markovian quantum state diffusion (NMQSD) equation \cite{DiosiNonMarkovianquantumstate1998, StrunzOpenSystemDynamics1999} by means of the well tested \ac{HOPS} approach \cite{SuessHierarchyStochasticPure2014, ZhangNonPerturbativeCalculationTwoDimensional2016, HartmannExactOpenQuantum2017, HartmannExactopenquantum2019}. 
To use this method for a \mbox{(sub-)} Ohmic environment the exact \ac{BCF} with algebraic decay has to be approximated by a sum of exponentials where the accuracy of the approximation can be controlled by the number of exponential terms.
Increasing the accuracy also increases the time to which the approximation follows the algebraic decay (see Fig. \ref{fig:OhmFit}).
Further details on how the \ac{HOPS} method has been applied successfully on (sub-) Ohmic environments can be found in Ref. \cite{HartmannExactOpenQuantum2017, HartmannExactopenquantum2019}.

Throughout this work we refer to the expression $c = \lambda_1 - \lambda_2 - \lambda_3 - \lambda_4$ as concurrence \cite{HillEntanglementPairQuantum1997, HorodeckiQuantumentanglement2009} where, strictly speaking, only the positive values of $c$ quantify entanglement.
However, when comparing the dynamics of $c$ it seems more convenient to also show its negative values.
As usual, $\lambda_i$ are the decreasingly sorted eigenvalues of the matrix $R = \sqrt{\sqrt{\rho} \tilde \rho \sqrt{\rho}}$ with $\tilde \rho = \sigma_y^A\sigma_y^B \rho^\ast \sigma_y^A\sigma_y^B$.
For numerical reason we calculate the eigenvalues $\lambda_i$ from the eigenvalues $a_i$ of the matrix $\rho \tilde \rho$ via $\lambda_i = \sqrt{a_i}$.

\subsection{Exact Dynamics}

To examine the properties of the entanglement generation we choose as initial condition the two-qubit product state $\ket{\psi_0} = \ket{\uparrow\uparrow}$ and a zero temperature environment where $\ket{\uparrow}$ ($\ket{\downarrow}$) is the eigenvector of $\sigma_z$ with eigenvalue $+1$ ($-1$).
Note, for the unbiased single qubit Hamiltonian ($\epsilon \sigma_z$ with $\epsilon = 0$) considered here the initial state $\ket{\downarrow\downarrow}$ yields the same entanglement dynamics and is, thus, not considered explicitly.
Also, the symmetry of the resonant Hamiltonian \eqref{eqn:hamiltonian} ($\omega_A = \omega_B$) results in a decoherence free subspace spanned by the Bell-state $\ket{\Phi_-} \sim \ket{\uparrow\downarrow} - \ket{\downarrow\uparrow}$.
Therefore, since the product states $\ket{\uparrow\downarrow}$ and $\ket{\downarrow\uparrow}$ contain a contribution from the decoherence free subspace, it is not too surprising that entanglement is generated \cite{SahrapourTunnelingDecoherenceEntanglement2013} if they serve as initial condition.
Consequently, they are not considered here but will be discussed elsewhere for detuned qubits.

\begin{figure}[tb]
  \includegraphics[width = \columnwidth]{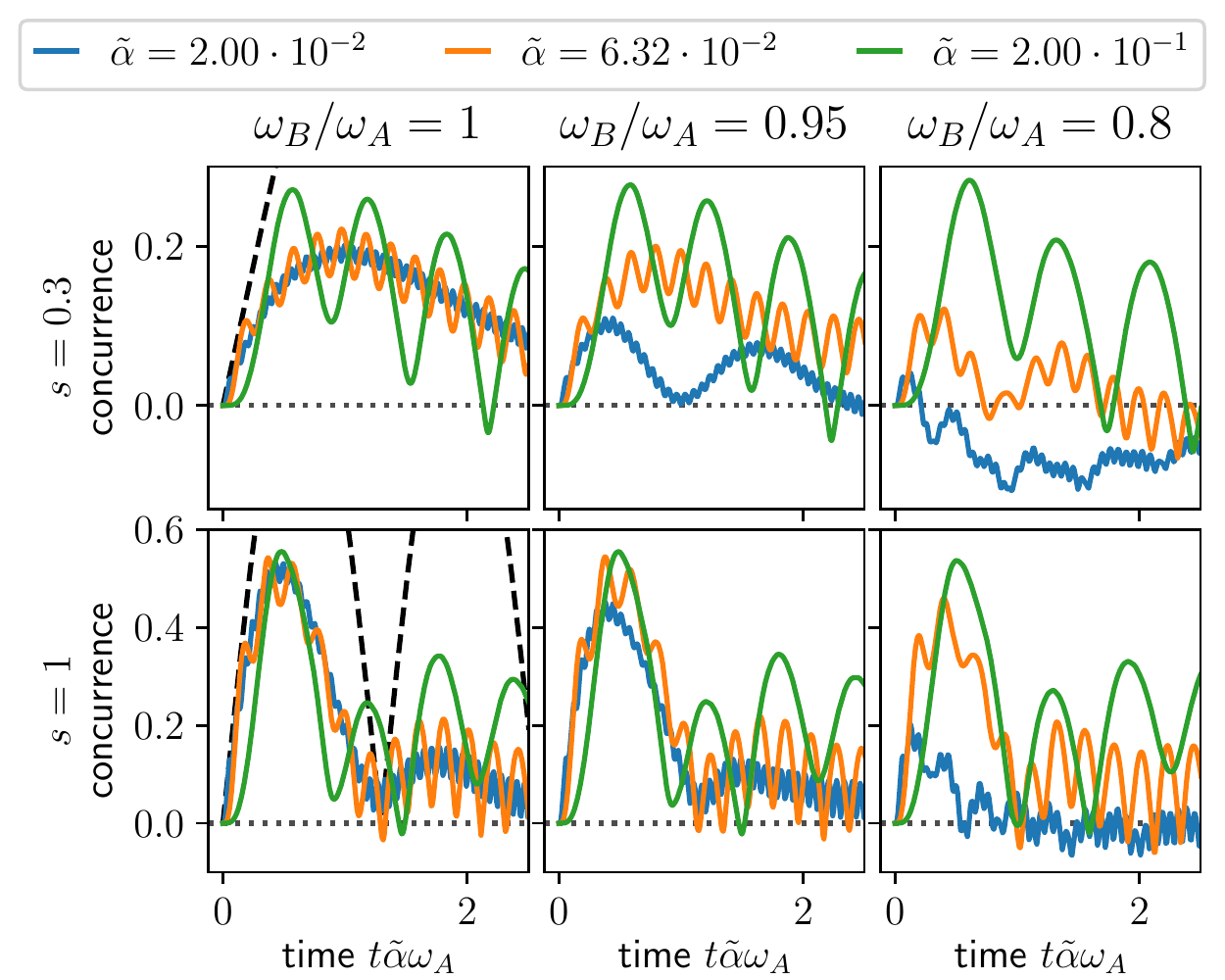}
  \caption{The dynamics of the two-qubit entanglement quantified by concurrence is shown for a sub-Ohmic ($s=0.3$) and Ohmic environment with $\omega_c = 10\omega_A$ and various coupling strengths.
  By keeping $\tilde \alpha = \alpha (\omega_c / \omega_A)^{1-s}$ constant the relaxation takes place on the same time scale for different parameters $s$.
  In addition different detuning parameters $\omega_B / \omega_A$ are considered.
  The dynamics shown here was obtained by the \ac{HOPS} method with an 5-term-exponential approximation of the \ac{BCF} which obeys an absolute error smaller than $10^{-3}$. The dynamics obtained using an even more accurate representation is indistinguishable with respect to the shown plots.
  In the same manner the convergence with respect to the hierarchy depth $k_\mathrm{max}$ and the number of samples $N$ has been checked. 
  In this sense the dynamics shown here is exact.
  The graphs shown were obtained by using a simple simplex truncation with $k_\mathrm{max} = 2$ and $N=4096$ samples for $\tilde \alpha=2.00 \cdot 10^{-2}$, $k_\mathrm{max} = 4$ and $N=16384$ for $\tilde \alpha=6.32 \cdot 10^{-2}$ and $k_\mathrm{max} = 4$ and $N=16384$ for $\tilde \alpha=2.00 \cdot 10^{-1}$.
  }
  \label{fig:HOPSDyn}
\end{figure}

As shown in Fig. \ref{fig:HOPSDyn}, for the resonant case $\omega_A = \omega_B$ a substantial amount of entanglement builds up initially (referred to the rescaled time $t \omega_A \alpha (\omega_c / \omega_A)^{1-s}$) for any weak to intermediate coupling strength as well as a sub-Ohmic and Ohmic environment.
Subsequently, the entanglement diminishes.
This kind of behavior can easily be explained by the interplay between the bath induced unitary interaction \cite{McCutcheonLonglivedspinentanglement2009, HartmannAccuracyAssessmentPerturbative2020} and the relaxation of the system \cite{CarvalhoEntanglementDynamicsDecoherence2007, YuSuddenDeathEntanglement2009} (see also Sec. \ref{sec:RWA}).

\begin{figure}[tb]
  \includegraphics[width = \columnwidth]{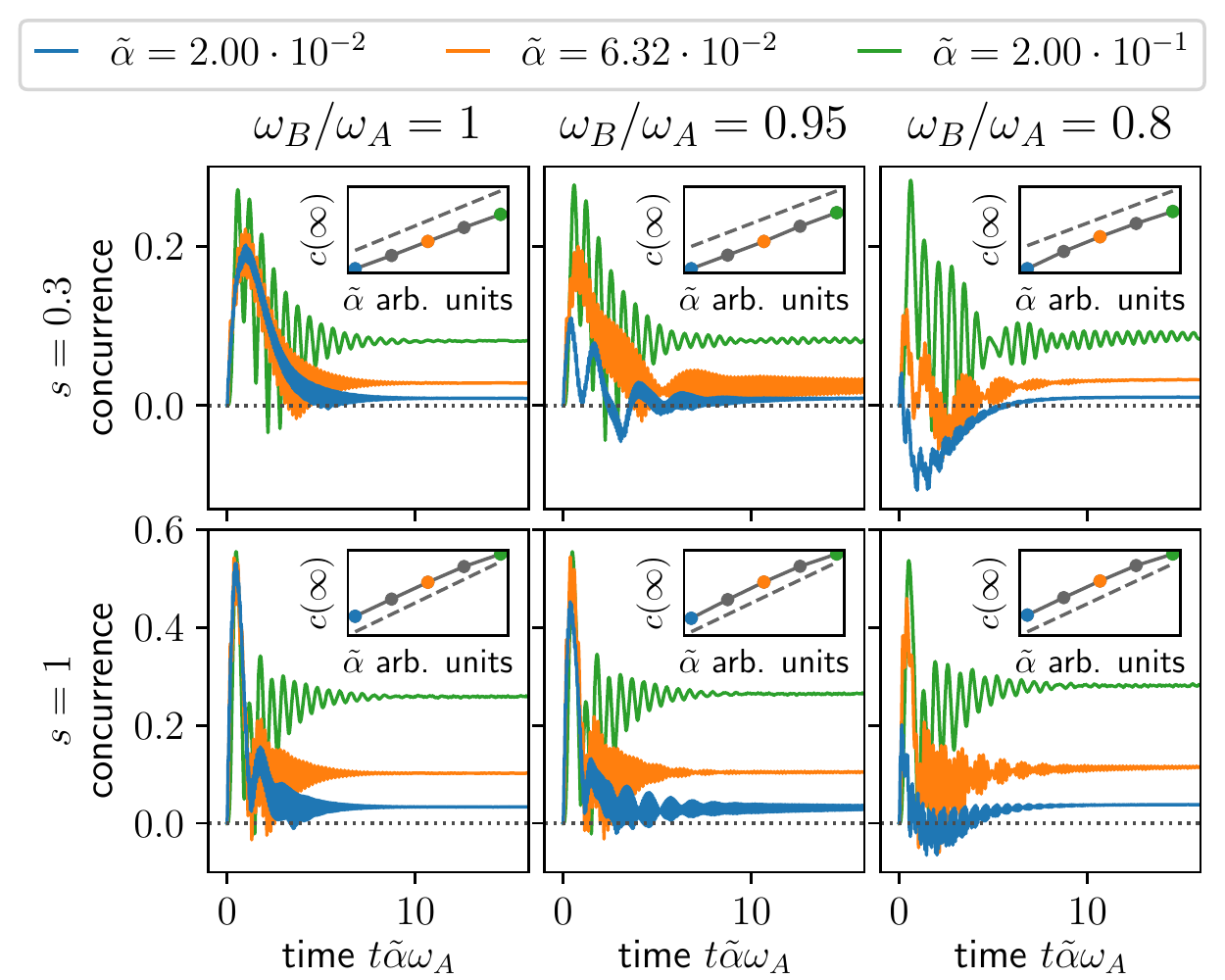}
  \caption{The entanglement dynamics is shown with special emphasize on the asymptotic behavior (same parameters as in Fig. \ref{fig:HOPSDyn}).  The insets show the linear dependence of the asymptotic entanglement as a function of the coupling strength (double logarithmic plot).}
  \label{fig:HOPSDynLong}
\end{figure}

Our calculations show that for the slightly detuned scenario $\omega_B / \omega_A = 0.95$ the entanglement evolves very similar to the resonant case.
As expected, differences are most prominent in the weak coupling regime.
For a larger detuning $\omega_B / \omega_A = 0.8$ such differences become more significant.

Notably, our exact results reveal that in any case the asymptotic entanglement does not vanish (see Fig. \ref{fig:HOPSDynLong}).
A result which cannot be obtained correctly by any of the perturbative approaches considered in the following.
Further, the asymptotic value is rarely influenced by the detuning.
This is particularly remarkable in the weak coupling regime where the initial dynamics is very sensitive to the detuning.

\subsection{Perturbative Master Equations}

In addition to the exact calculations the properties of the dynamics obtained by perturbative master equations are of relevance, too.
The motivation is given by the rather simple structure of such master equations which allows to gain insights in the relevant mechanisms.
However, the suitability of these perturbative approaches has to be checked since an error estimation from within the perturbative formalism seems unfeasible.
With the help of the exact dynamics obtained by the \ac{HOPS} method we examine the applicability of the \ac{QOME} (Born-Markov approximation and \ac{RWA}) \cite{BreuerTheoryOpenQuantum2007, KryszewskiMasterequationtutorial2008}, its variation with only a \ac{PRWA} \cite{VogtStochasticBlochRedfieldtheory2013, JeskeBlochRedfieldequationsmodeling2015, TscherbulPartialsecularBlochRedfield2015}, the Redfield equation (RFE) (no \ac{RWA} at all) \cite{RedfieldTheoryRelaxationProcesses1957}, the very recent \ac{GAME} (\ac{GKSL} kind equation based on the \ac{RFE}) \cite{DavidovicCompletelyPositiveSimple2020} and the coarse-graining approach \cite{SchallerPreservationPositivityDynamical2008, BenattiEntanglingtwounequal2010, MajenzCoarseGrainingCan2013}, in the context of two independent qubits coupled to a common environment with sub-Ohmic \ac{SD}.

\subsubsection{The Rotating Wave Approximation}

\label{sec:RWA}

The great success of the \ac{QOME} (Born-Markov approximation and full \ac{RWA}, well known \ac{GKSL} form, see Ref. \cite{BreuerTheoryOpenQuantum2007, KryszewskiMasterequationtutorial2008, HartmannAccuracyAssessmentPerturbative2020}) to describe the dynamics of a single qubit encourages the use of the same formalism for two qubits also.
Since the \ac{QOME} is of \ac{GKSL} form positivity of the reduced state is assured for all times and any initial condition.
For the resonant case $\omega_A = \omega_B$ it is easily seen that the unitary contribution of the \ac{QOME} (Lamb shift term) \cite{BenattiEntanglingoscillatorsenvironment2006, McCutcheonLonglivedspinentanglement2009, HartmannAccuracyAssessmentPerturbative2020}
\begin{multline}
  H_\mathrm{Lamb}  = H_\mathrm{local} \\
  + \frac{S(\omega_A) + S(-\omega_A)}{8} \left(\sigma_z^A \sigma_z^B + \sigma_y^A\sigma_y^B \right)
\end{multline}
couples the qubits such that entanglement builds up \cite{LendiDaviestheoryreservoirinduced2006, SolenovExchangeinteractionentanglement2007, McCutcheonLonglivedspinentanglement2009, SahrapourTunnelingDecoherenceEntanglement2013} (see dashed graphs in Fig. \ref{fig:HOPSDyn}).
Here $S$ denotes the imaginary part of the half sided Fourier transform $F(\omega)$ of the zero temperature \ac{BCF} $\alpha_\mathrm{bcf}$ which is defined by means of the \ac{SD} $J(\omega)$
\begin{equation}
\begin{gathered}
  \alpha_\mathrm{bcf}(\tau) = \frac{1}{\pi} \il{0}{\infty}{\omega} J(\omega)^{-\imag\omega\tau} \\
  F(\omega) = \il{0}{\infty}{\tau} \alpha_\mathrm{bcf}(\tau)^{\imag\omega\tau} = J(\omega) + \imag S(\omega) .
  \label{eqn:halfsidedFT}
\end{gathered}
\end{equation}
Since the local contribution $H_\mathrm{local}$ acts on a single qubit only it does not influence the entanglement dynamics.

\begin{figure}[tb]
  \includegraphics[width = \columnwidth]{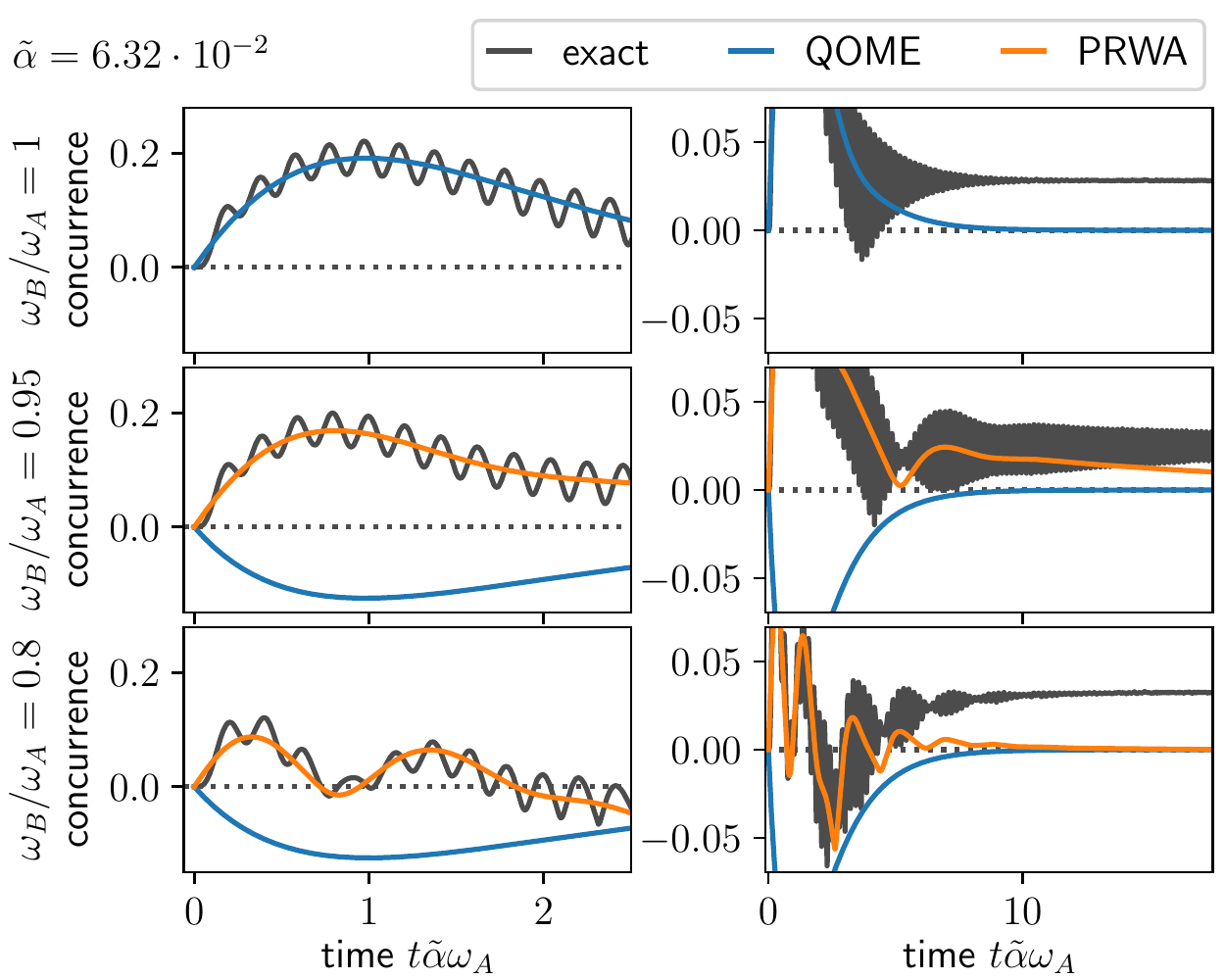}
  \caption{The entanglement dynamics obtained from the \ac{QOME} and \ac{PRWA} approach is shown together with the exact dynamics of the \ac{HOPS} method for a sub-Ohmic environment with $s=0.3$, $\omega_c=10\omega_A$ and $\tilde \alpha = 6.32 \cdot 10^{-2}$. The short time dynamics (left column) and long time dynamics (right column) is shown for different values of detuning $\omega_B/\omega_A$ (rows). Note, the \ac{QOME} and the \ac{PRWA} coincide in the resonant case. Whereas the \ac{QOME} is not able to predict the entanglement generation for detuned qubits, the \ac{PRWA} approach agrees well with the exact dynamics (except for the superimposed oscillations). Both master equations yield a vanishing asymptotic entanglement whereas the exact value remains finite.}
  \label{fig:RWA}
\end{figure}

Importantly, as of the \ac{RWA} involved in the derivation of the \ac{QOME}, for detuned qubits $\omega_A \neq \omega_B$ the Lamb shift term consists of the local part only and, thus, entanglement generation is not featured by the \ac{QOME} \cite{BenattiEntanglingtwounequal2010}.
However, as shown in Fig. \ref{fig:RWA} the exact dynamics obtained from the \ac{HOPS} method reveals that for slightly detuned qubits entanglement is generated in a very similar manner as in the resonant case (see also Ref. \cite{Flemingrotatingwaveapproximationconsistency2010, MaEntanglementdynamicstwo2012, EasthamBathinducedcoherencesecular2016, HartmannAccuracyAssessmentPerturbative2020} for more detailed discussions on the severe consequences of the \ac{RWA} on bipartite correlations).
Assuming that the \ac{QOME} provides a suitable approximation in some weak coupling regime it seems contradictory that an infinitesimal change of the system parameters results in a significant change of the dynamics.
The inconsistency can be resolved by recalling that a small detuning introduces a new very slow system time scale.
Only for a decay time scale (inverse of the coupling strength) much larger than the system time scale the \ac{RWA} and, thus, the \ac{QOME} is applicable.
This means that the applicability of the \ac{QOME} crucially depends on the system parameters.
Only in the limit of zero coupling (in combination with a rescaled time, the so called scaling or van Hove limit) the \ac{QOME} becomes exact \cite{daviesMarkovianMasterEquations1976}.
This means that for a sufficiently weak coupling strength the entanglement generation will vanish for any detuning, but remain in the resonant case.
This behavior can, to some extent, be seen from the exact dynamics shown in Fig. \ref{fig:HOPSDyn}.

After having pointed out the severe problems in the detuned case arising from the \ac{RWA} we consider perturbative approaches which aim to circumvent the \ac{RWA}.
A first one, very similar to the \ac{QOME}, makes use of the \ac{RWA} only partially \cite{VogtStochasticBlochRedfieldtheory2013, JeskeBlochRedfieldequationsmodeling2015, TscherbulPartialsecularBlochRedfield2015} which results in a master equation of \ac{GKSL} kind also.
The specific expression of the master equation for the model considered here can be found in Ref. \cite{HartmannAccuracyAssessmentPerturbative2020}.
The master equation using the \ac{PRWA} is constructed from the same non-local Lindblad operators occurring in the \ac{QOME} for resonant qubits.
Henceforth, the Lamb shift term is non-local and features entanglement generation also in the detuned case.

The initial entanglement dynamics obtained from the \ac{PRWA} master equation agrees quantitatively, up to some fast oscillation, with the exact dynamics obtained by the \ac{HOPS} method (see Fig. \ref{fig:RWA}).
The fast superimposed oscillation are not captured because the \ac{PRWA} still neglects some secular terms.
Keeping all secular terms yields the \ac{RFE} considered next.
Both, the \ac{QOME} and the \ac{PRWA} approach yield a vanishing asymptotic entanglement which is in disagreement with the exact dynamics.

\subsubsection{The Redfield Equation}

\begin{figure}[tb]
  \includegraphics[width = \columnwidth]{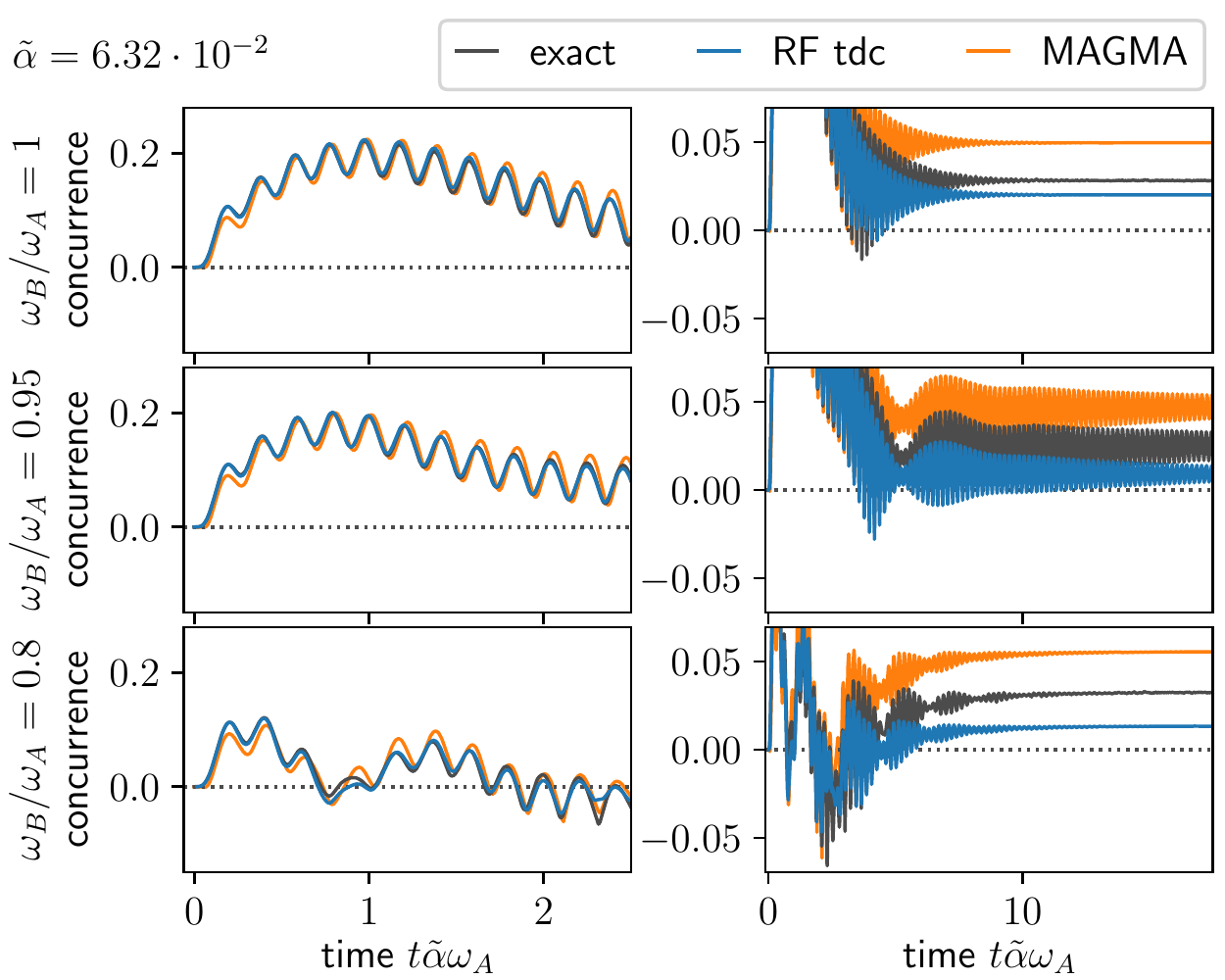}
  \caption{The entanglement dynamics obtained from the \ac{RFE} ($\rho \rightarrow \sqrt{\rho\rho^\dagger}$), the \ac{GAME} master equation and exact dynamics is shown for the same parameters as in Fig. \ref{fig:RWA}. For all three detuning parameters the \ac{RFE} matches the exact dynamics very well over a significant period of time. For larger times deviations become apparent. Despite the fact that the asymptotic entanglement from the \ac{RFE} does not vanish it disagrees with the exact value. Although slightly less accurate the \ac{GAME} yields comparable results to the \ac{RFE}.}
  \label{fig:RF}
\end{figure}

Motivated by the detailed accuracy assessment of perturbative master equations \cite{HartmannAccuracyAssessmentPerturbative2020} we consider not only all secular terms but also use the \ac{RFE} with time dependent coefficients which was shown to be the most accurate method.
As shown in Fig. \ref{fig:RF} the high degree of accuracy also holds for the sub-Ohmic environment considered here.
The entanglement dynamics obtained from the \ac{RFE} with time dependent coefficients matches very well the exact dynamics.
Even the fast oscillations are recovered.

\begin{figure}[tb]
  \includegraphics[width = \columnwidth]{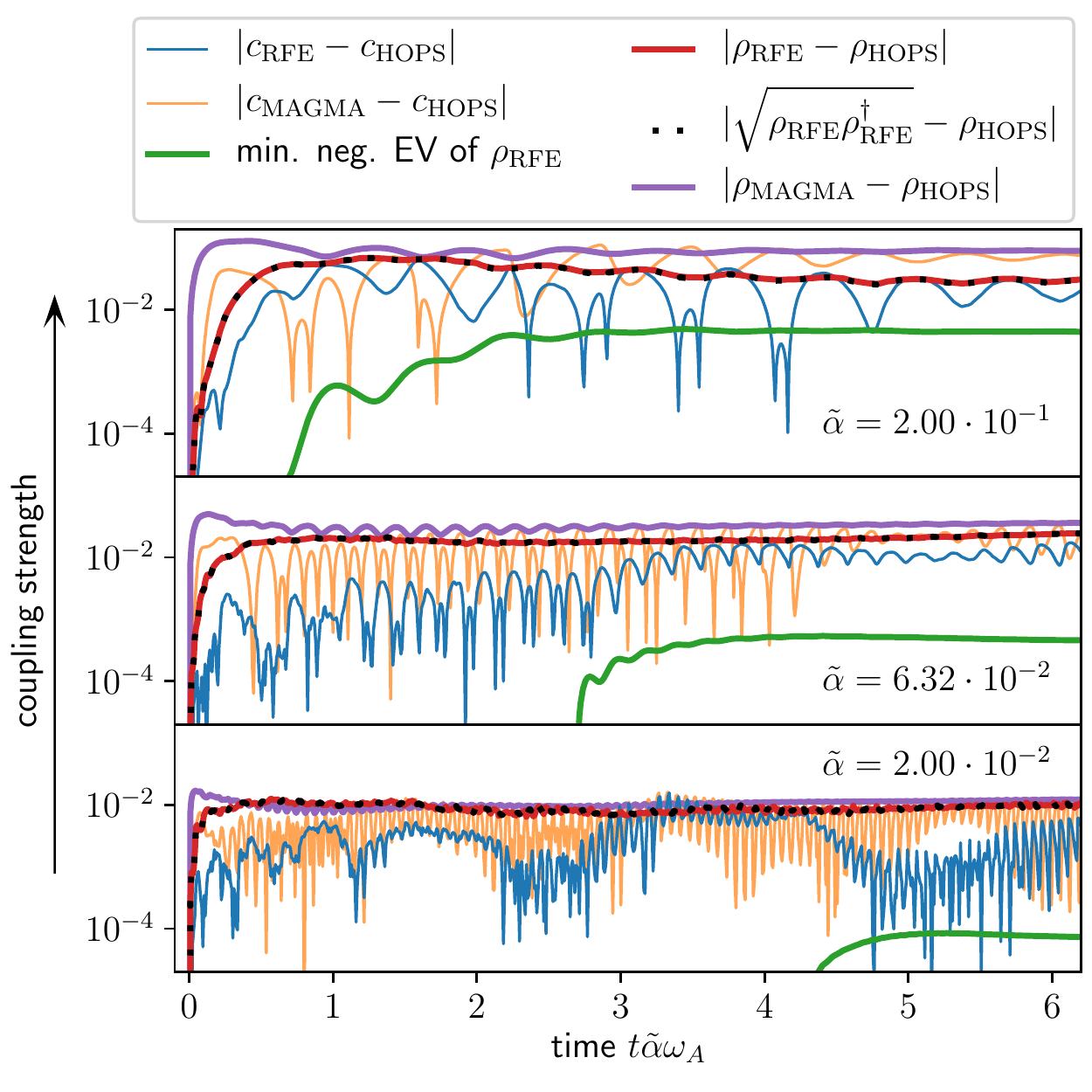}
  \caption{To justify the use of the positive matrix $\sqrt{\rho \rho^\dagger}$ used to calculate the concurrence for the \ac{RFE} the deviation of $\rho$ (red line) and $\sqrt{\rho \rho^\dagger}$ (black dots) from the exact state is shown using the Hilbert-Schmidt-norm. The deviations are nearly indistinguishable which means that $\sqrt{\rho \rho^\dagger}$ approximates the exact state as good as $\rho$. This is in accordance with the magnitude of the smallest negative eigenvalue of $\rho$ which is significantly smaller than the deviation. In addition the absolute difference between the exact and the \ac{RFE} concurrence is shown (blue line). The error of the concurrence exceeds the error of the state only marginally. 
  Furthermore, the error of the concurrence (yellow line) as well as the density matrix (purple line) is plotted also for the \ac{GAME}. As of the additional approximation involved the approach is slightly less accurate compared to the \ac{RFE}. Nonetheless, the \ac{GAME} outperforms the other \ac{GKSL} kind equations considerably.}
  \label{fig:RF_diff}
\end{figure}

Small positivity violations of the reduced dynamics occur already in a regime where the accuracy is still acceptable (see Fig. \ref{fig:RF_diff}).
Consequently, using the positivity violation as an indicator for the break down of the weak coupling assumption, as proposed for a Lorentzian environment with an exponentially decaying \ac{BCF} \cite{HartmannAccuracyAssessmentPerturbative2020}, is not that simple for a sub-Ohmic \ac{SD} with a slow algebraic decay of the \ac{BCF}.
The positivity violation is of particular relevance here because the concurrence is not defined for non-positive $\rho\tilde\rho$ even though $\rho$ approximates the exact state suitably well.
However, since the positivity violation is of the order of the perturbative error the matrix $\sqrt{\rho \rho^\dagger}$ is consistent with the degree of perturbation (see Fig. \ref{fig:RF_diff}) while being positive by construction.
This justifies the calculation of the concurrence using the positive matrix $\sqrt{\rho \rho^\dagger}$ obtained from the reduced dynamics of the \ac{RFE}.

\subsubsection{The Geometric-Arithmetic Master Equation (GAME)}

In a recent publication the failure to quantify entanglement of a non-positive state in an approximative sense has been addressed as well \cite{DavidovicCompletelyPositiveSimple2020}.
The proposed \ac{GAME} modifies the \ac{RFE} such that it becomes a master equation of \ac{GKSL} type.
Crucially, the environmentally induced unitary influence on the system is identified from the \ac{RFE} before the approximation is applied.
The justification of \ac{GAME} is based on the relaxation time scale in the interaction picture, roughly given by the inverse of the coupling strength, and properties of the \ac{SD} only.
In contrast to the \ac{QOME} the particular spectrum of the system Hamiltonian is irrelevant.
It is, thus, expected that the entanglement generation is well captured for any detuning of the qubits.
As shown in Fig. \ref{fig:RF} this is indeed the case.
The \ac{GAME} master equation mimics even the fast oscillations of the entanglement dynamics, however, slightly less accurate compared to the positive state constructed from the \ac{RFE} dynamics (see also Fig. \ref{fig:RF_diff} which shows the error).
Since the \ac{GAME} master equation is based on the \ac{RFE} it is consistent that the asymptotic entanglement deviates from the exact value on the same scale as the \ac{RFE} result (see Fig. \ref{fig:RF}), although the value is overestimated.
This allows to conclude that the positivity issues related to the evaluation of the entanglement can be cured slightly more accurate using the Redfield formalism with $\rho_{\mathrm{aprx}} := \sqrt{\rho \rho^\dagger}$ than applying the additional \ac{GAME}.

\subsubsection{The Coarse-Graining Master Equations}

\begin{figure}[tb]
  \includegraphics[width = \columnwidth]{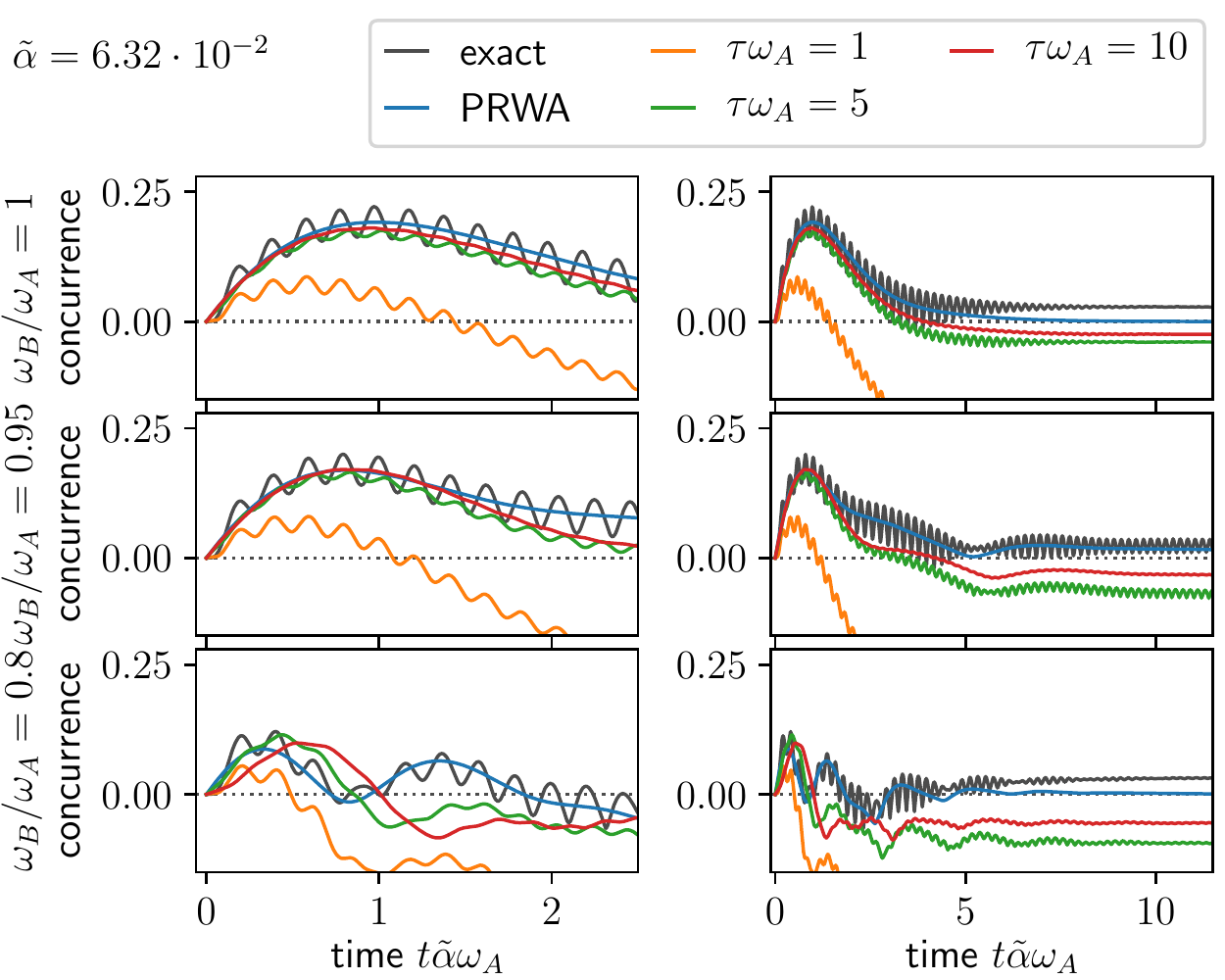}
  \caption{The entanglement dynamics of the \ac{CGME} is shown for comparison (same environmental parameters as in Fig. \ref{fig:RWA}). Since the \ac{QOME} is suitable in the resonant case it is not surprising that for large coarse-graining parameters $\tau$ the \ac{CGME} yields similar results. For the detuned case the entanglement generation is still visible, in contrast to the \ac{QOME}. However, the \ac{PRWA} agrees better with the exact results than the \ac{CGME} (see the text for a discussion about the time scale separation required by the \ac{CGME}).}
  \label{fig:CG}
\end{figure}

An alternative path towards a master equation for the microscopic model which assures positive dynamics is the so-called coarse-graining procedure \cite{SchallerPreservationPositivityDynamical2008, BenattiEntanglingtwounequal2010, MajenzCoarseGrainingCan2013}.
It has been proposed precisely with the aim to overcome the limitations due to the \ac{RWA}.
In particular for the case of two detuned qubits entanglement generation has been shown using such a \ac{CGME} \cite{BenattiEntanglingtwounequal2010}.
We confirm this result for a common sub-Ohmic environment (see Fig. \ref{fig:CG}).
However, the quantitative comparison for a sub-Ohmic environment with $s=0.3$ and $\omega_c = 10\omega_A$ shows that the \ac{CGME} does not reach the accuracy of the \ac{PRWA} nor the \ac{GAME}.
To understand the deviation of the \ac{CGME} recall that the condition for its applicability reads $\tau_\env \ll \tau \ll \tau_\mathrm{ind.}$ \cite{HartmannAccuracyAssessmentPerturbative2020}.
The environmentally induced timescale for the system dynamics scales with the inverse coupling strength.
From the exact dynamics in Fig. \ref{fig:CG} its follows $\tau_\mathrm{ind} \approx 0.2 (\tilde \alpha \omega_A)^{-1} \approx 3 \omega_A^{-1}$.
Although the decay time of the \ac{BCF} scales with the cutoff frequencies $\tau_\env \sim \omega_c^{-1}$ the particular time at with the \ac{BCF} has decayed over, for example, one order of magnitude is $\tau_{\env,1} \approx 10^{1/(s+1)} \omega_c^{-1}$.
For the example considered here it follows $\tau_{\env,1} \approx 0.6\omega_A^{-1}$.
Obviously, a clear separation of the three timescales is not justified.

\subsubsection{Summary}

As expected, the approach with the least approximations, namely the \ac{RFE} with time dependent coefficients, yields the most accurate results for the entanglement dynamics of two qubits coupled to a common sub-Ohmic environment.
Positivity violations can consistently be healed by using the positive matrix $\sqrt{\rho \rho^\dagger}$ to estimate the concurrence.
The \ac{PRWA}, which coincides with the \ac{QOME} in the resonant case, yields the correct overall dynamics while missing faster oscillations.
The \ac{CGME} is less appropriate for the set of parameters considered here.
Note that the recent \ac{GAME} outperforms the other \ac{GKSL}-type master equations in terms of accuracy.
However, as of the simple structure of the \ac{PRWA}, we use this approach together with the exact results to illuminate the influence of the counterterm in the next section.
The reasoning could have been done equally well based on the \ac{GAME} since due to the weak coupling both approaches yield nearly indistinguishable dynamics.

\section{Influence of the Counterterm}

To recapitulate the motivation for the counterterm -- compensation of the environmentally induced unitary effect on the system dynamics -- a particle with position $q$ and momentum $p$ moving in a potential $V(q)$ and coupled to a set of harmonic oscillators with position (momentum) $x_\lambda$ ($p_\lambda$) is considered.
The linear coupling $-\sum_\lambda F_\lambda(q) x_\lambda$ to the environmental modes contributes to the potential affecting the particle.
It has been argued that in the adiabatic regime (instantaneous adjustment of the environmental modes to the particle position) the effective potential becomes \cite{WeissQuantumDissipativeSystems2008}
\begin{equation}
  V_\mathrm{eff}(q) = V(q) - \sum_\lambda \frac{F_\lambda(q)^2}{2 m_\lambda \omega_\lambda^2} \quad .
\end{equation}
For a mode independent coupling operator $F_\lambda(q) = c_\lambda F(q)$, as in Eq. \eqref{eqn:hamiltonian}, the Hamiltonian
\begin{multline}
  H = \frac{p^2}{2M} + V(q) + \\
  \sum_\lambda \left[\frac{p_\lambda^2}{2m_\lambda} + \frac{m_\lambda \omega_\lambda^2}{2}\left(x_\lambda - \frac{c_\lambda}{m_\lambda \omega_\lambda^2}F(q)\right)^2 \right]
  \label{eqn:hamiltonian_with_CT}
\end{multline}
compensates that change of the potential. 
It is assumed that the same holds true when casting the above Hamiltonian to the more abstract form of Eq. \eqref{eqn:hamiltonian}.
In particular the relations $g_\lambda \sqrt{2m_\lambda \omega_\lambda} = c_\lambda$ and $L = -F(q)$ yield for the so-called counterterm
\begin{multline}
H_\mathrm{c} = \sum_\lambda \frac{c_\lambda^2 F(q)^2}{2 m_\lambda \omega_\lambda^2}  = \sum_\lambda \frac{g_\lambda^2}{\omega_\lambda} L^2 \\
             = \frac{1}{\pi}\il{0}{\infty}{\omega} \frac{J(\omega)}{\omega} L^2
\end{multline}
which is expected to renormalize the system Hamiltonian.
For the class of (sub-) Ohmic SD, as given in Eq. \eqref{eqn:hamiltonian}, the counterterm evaluates to $H_c = \frac{\alpha \omega_c}{2} \Gamma(s) L^2$.

\begin{figure*}[tb]
  \includegraphics[width = \textwidth]{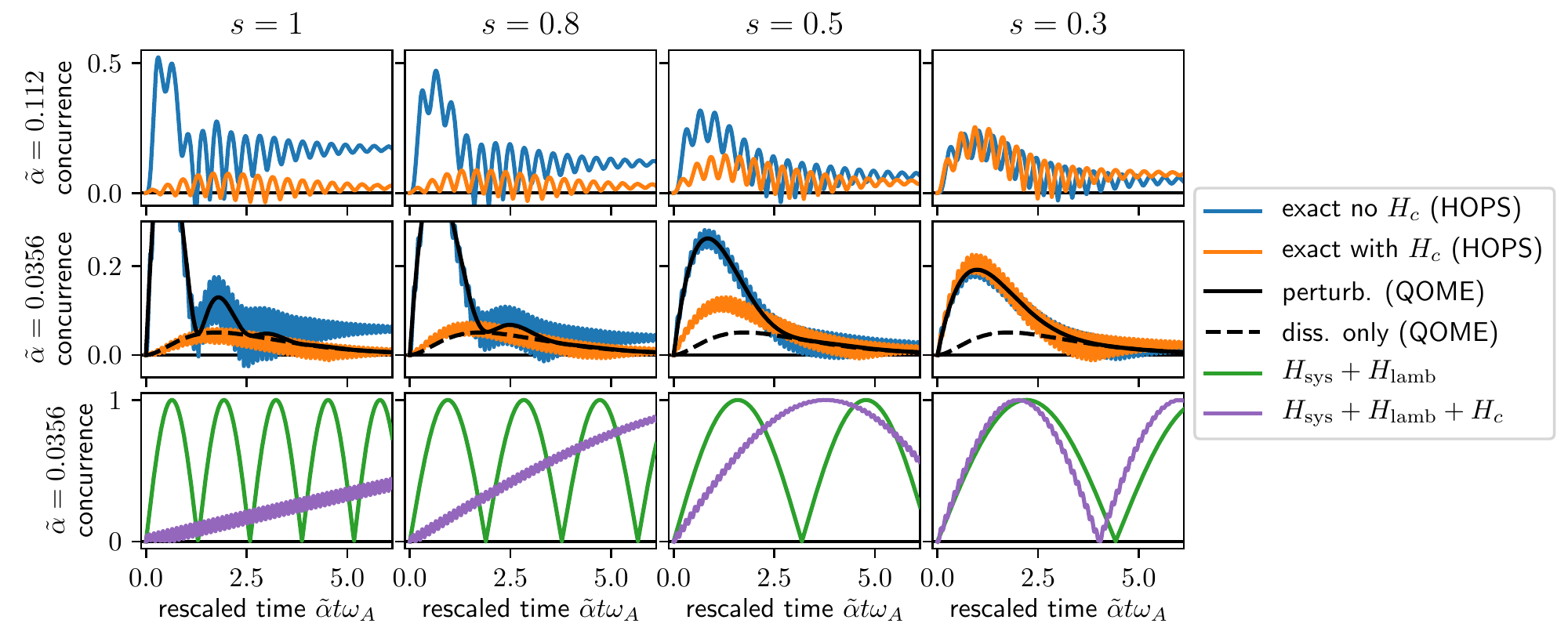}
  \caption{The influence of the counterterm $H_c$ on the entanglement dynamics for two resonant qubits is shown (blue lines: exact dynamics without $H_c$, orange lines exact dynamics including $H_c$, $\omega_c / \omega_A = 10$). For weak and intermediate coupling strength the influence is evidently visible in the Ohmic case, however, diminishes in the deep sub-Ohmic regime. 
  It seems obvious that if the counterterm compensates the Lamb shift Hamiltonian in the weak coupling regime the \ac{QOME} with dissipator only (dashed line) should yield a similar behavior to the exact dynamics including the counter term (orange line).
  The second row shows that this assertion holds in the Ohmic case but fails spectacularly for small values like $s=0.3$.
  This behavior is further confirmed by looking at the unitary dynamics only (third row).
  The system Hamiltonian in combination with the Lamb shift results in generation of entanglement (green line) which is reflected by the short time dynamics of the exact dynamics without the counterterm (blue line).
  Adding the counterterm (purple line) the entanglement buildup is slowed down significantly in the Ohmic case.
  Therefore, on the timescale of the relaxation the counterterm cancels the effect of the Lamb shift term.
  However, the speed of the entanglement generation increases with decreasing $s$ (purple lines) which manifests that the counterterm cancels the Lamb shift term less effectively for small $s$.}
  \label{fig:counterterm}
\end{figure*}

\begin{figure}[tb]
  \includegraphics[width = \columnwidth]{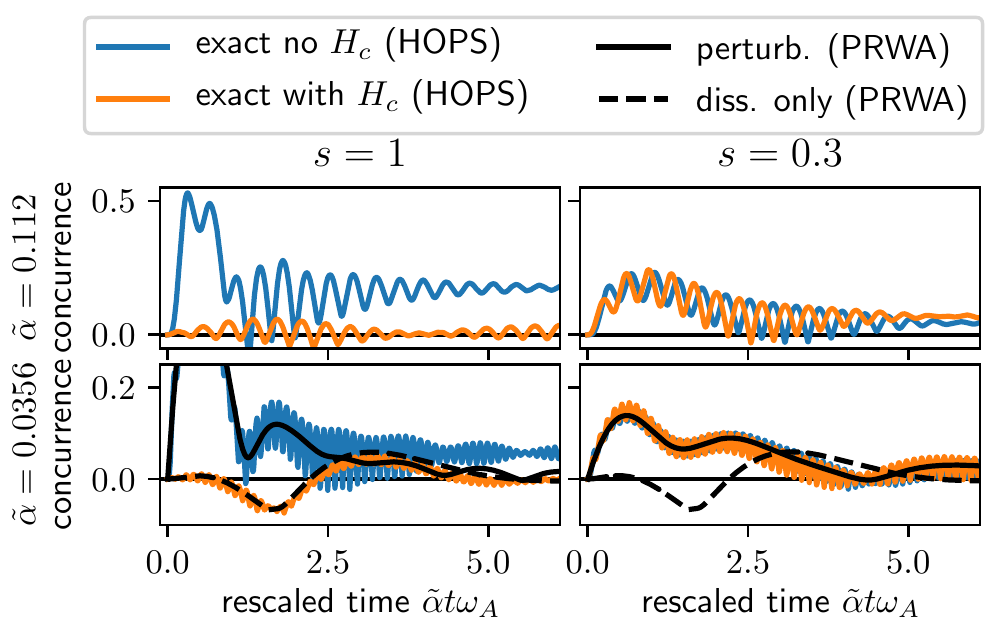}
  \caption{For slightly detuned qubits ($\omega_B / \omega_A = 0.95$) and a cutoff frequency $\omega_c / \omega_A = 10$ the entanglement dynamics under the influence of the counterterm is shown (same coloring as in Fig. \ref{fig:counterterm}). 
  As of the detuning the \ac{PRWA} master equation is used to obtain the perturbative dynamics. 
  Similar to the case of resonant qubits, the counterterm approximately cancels the Lamb shift term in the Ohmic case. As shown in the lower left panel, the exact dynamics including the counter term (orange line) agrees with the dynamics under the action of the \ac{PRWA} dissipator only (dashed black line).
  However, this expected behavior is not observed for a sub-Ohmic \ac{SD} with small $s$. 
  Including the counterterm rarely affects the entanglement dynamics (blue and orange lines, lower right panel).
  The influence of the cutoff frequency is shown in Fig. \ref{fig:large_wc_small_s_1_0_3}.}
  \label{fig:counterterm_db_0_95}
\end{figure}

It will be shown in the following that for two non-interacting qubits it is primarily the environmentally induced unitary interaction which generates entanglement.
Therefore, including the counterterm is expected to suppress the generation of entanglement.
As an example, its influence on the entanglement dynamics is shown in Fig. \ref{fig:counterterm} (resonant qubits) and Fig. \ref{fig:counterterm_db_0_95} (detuned qubits).
Two main features are observed.
First, with decreasing $s$ (going from the Ohmic to the deep sub-Ohmic regime) the impact of the counterterm becomes less, independent of the coupling strength (see first and second row in Fig. \ref{fig:counterterm}).
Second, for weak coupling and $s=1$ the entanglement dynamics including the counterterm is very well mimicked by the action of the dissipator only of the \ac{PRWA} master equation (which coincides with the \ac{QOME} in the resonant case).
This shows that the counterterm does compensate the induced unitary interaction up to a certain level.
However, for $s=0.3$ the exact dynamics, hardly influenced by $H_c$, does not agree with the dissipative dynamics in the perturbative regime.
Consequently the above assertion about the effect of the counterterm does not hold true in the deep sub-Ohmic regime.

To understand this behavior we investigate the dynamics in the weak coupling regime.
In that regime the \ac{RFE} is known to be very accurate \cite{HartmannAccuracyAssessmentPerturbative2020}.
Using asymptotic coefficients it reads
\begin{equation}
  \dot \rho = -\imag [H_\mathrm{sys}, \rho] + \sum_{i} \left( F(\omega_i) [L_{\omega_i} \rho, L] + \hc \right)
\end{equation}
where $L_{\omega_i} = \sum_{\{\epsilon, \epsilon' | \epsilon'-\epsilon = \omega_i\}} \ket{\epsilon}\bra{\epsilon}L \ket{\epsilon'}\bra{\epsilon'}$ is the decomposition of the coupling operator by means of projectors of eigenstates of the system Hamiltonian $H_\mathrm{sys} \ket{\epsilon} = \epsilon \ket{\epsilon}$ and the coefficients $F(\w_i)$ are the values of the half-sided Fourier transform of the \ac{BCF} [Eq. \eqref{eqn:halfsidedFT}]. 
It is well known from several perturbative treatments that the real part of $F(\omega) = J(\omega) + \imag S(\omega)$ accounts for the dissipative dynamics whereas the imaginary part determines the unitary contribution.
Under the assumption that the particular values $S(\omega_i)$ can be approximated by $S(0)$ (recall $\omega_i$ takes the value of all possible transition frequencies of the system Hamiltonian) the corresponding contribution in the \ac{RFE}
becomes
\begin{equation}
    \imag \sum_i \left(S(\omega_i) [L_{\omega_i} \rho(t), L] + \hc\right) \approx - \imag S(0) [L^2, \rho(t)]
\end{equation}
where we have used that by definition $L = \sum_i L_{\omega_i}$.
For zero temperature $S(\omega)$ is related to the SD by
\begin{equation}
  S(\omega) = - \frac{1}{\pi} \mathcal{P} \il{0}{\infty}{\omega'} \frac{J(\omega')}{\omega' - \omega}
\end{equation}
where $\mathcal{P}$ denotes the Cauchy principal value.
We finally conclude that given $S(\omega_i) \approx S(0)$ the unitary contribution in the perturbative treatment (Lamb shift term) is approximately canceled by the counterterm
\begin{equation}
  H_\mathrm{Lamb} \approx S(0) L^2 = - \frac{1}{\pi} \il{0}{\infty}{\omega} \frac{J(\omega)}{\omega} L^2 = -H_c \; .
\end{equation}

\begin{figure}[tb]
  \includegraphics[width = \columnwidth]{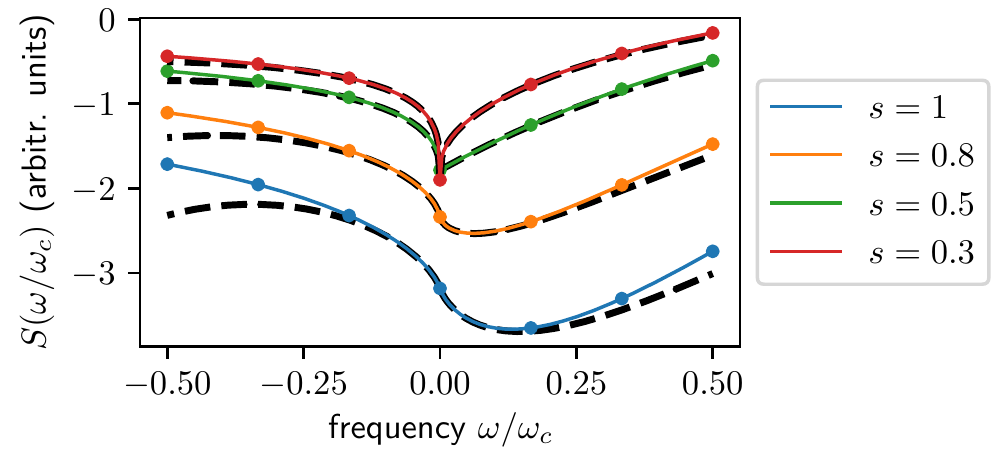}
  \caption{The behavior of the imaginary part of the half-sided Fourier transform of the BCF $S(\omega)$ is shown for different parameters $s$.
  (solid lines: exact analytic expression, dashed lines: expansion around zero, dots: numeric Fourier integral)}
  \label{fig:Sw}
\end{figure}

For the class of (sub-) Ohmic SD the validity of the assumption $S(\omega_i) \approx S(0)$ depends on the parameter $s$ in a very sensitive way.
To see that, it is convenient to expand the function $S(\omega)$ by introducing $x = \omega / \omega_c$.
Expressing $F(\omega)$ analytically 
\begin{multline}
  F(\omega) = \frac{\eta \Gamma(s+1)}{\pi} \frac{(-\imag \omega)^s}{\imag^{s+1}} e^{-x}\\
  \times \lim_{\epsilon \rightarrow 0} \Gamma\left(-s, -(x+\imag\epsilon)\right)
\end{multline}
(the limit assures the correct branch of the incomplete gamma function) and expanding the incomplete gamma function for small $x$ allows to write
\begin{equation}
    S(x) = - \frac{\alpha \Gamma(s)}{2} \omega_c e^{-x} f(s, x)
\end{equation}
where in the sub-Ohmic case $f$ expands to
\begin{equation}
\begin{aligned}
  f(s, x) & = 1 + s g(s, x) \Gamma(-s) |x|^s  + \frac{s x}{s-1} + \mathcal{O}(x^2) \\
  g(s, x) & = \cos(\pi s)\Theta(x) + \Theta(-x)
\end{aligned}    
\end{equation}
with $\Theta$ denoting the Heaviside step function.
For an Ohmic environment $f$ becomes
\begin{equation}
    f(1,x) = 1 - x \ln(|x|) + (1-e_\gamma)x + \mathcal{O}(x^2)
\end{equation}
where $e_\gamma$ denotes the Euler–Mascheroni constant.
The expansion shows that the lowest order behaves like $|x|^s$ in the sub-Ohmic and $x\log(|x|)$ in the Ohmic case (see Fig. \ref{fig:Sw} for example plots).
In addition, for positive $x$ the term $\sim|x|^s$ changes its sign at $s=0.5$ which results in a pointed minimum at $S(0)$ for $s < 0.5$.

It follows that the approximation $S(x\!=\!\omega_i/\omega_c) \approx S(0)$, obviously, becomes better when increasing $\omega_c$.
However, since $S(x\!=\!\omega_i/\omega_c)$ behaves like $\sim |x|^s$ around zero, for a fixed $\omega_c$ the approximation becomes significantly worse when changing from the Ohmic ($s=1$) to the deep sub-Ohmic ($s < 0.5$) regime.
This behavior explains that the cancellation of the Lamb shift Hamiltonian by the counterterm fails spectacularly for small $s$ and the cutoff frequency $\omega_c = 10\omega_A$ considered in the above examples.

In order to relate these considerations to the example dynamics shown above the dynamics of the unitary contributions are shown in Fig. \ref{fig:counterterm}, too (bottom row).
As expected, the pure influence of the Lamb shift on the system Hamiltonian results in oscillatory dynamics of the concurrence (green graph).
Adding the counterterm $H_c$ (purple graph) yields a significantly slower generation of entanglement for $s=1$.
The two contributions compensate, at least approximately, on the relaxation time scale.
However, as elucidated above for $s=0.3$ the effect of including $H_c$ is far from cancellation -- entanglement is generated even slightly faster.
This behavior is directly captured by the exact entanglement dynamics of the two-spin-boson model calculated using the HOPS method (Fig. \ref{fig:counterterm} and Fig. \ref{fig:counterterm_db_0_95}).

\begin{figure}[t!]
  \includegraphics[width = \columnwidth]{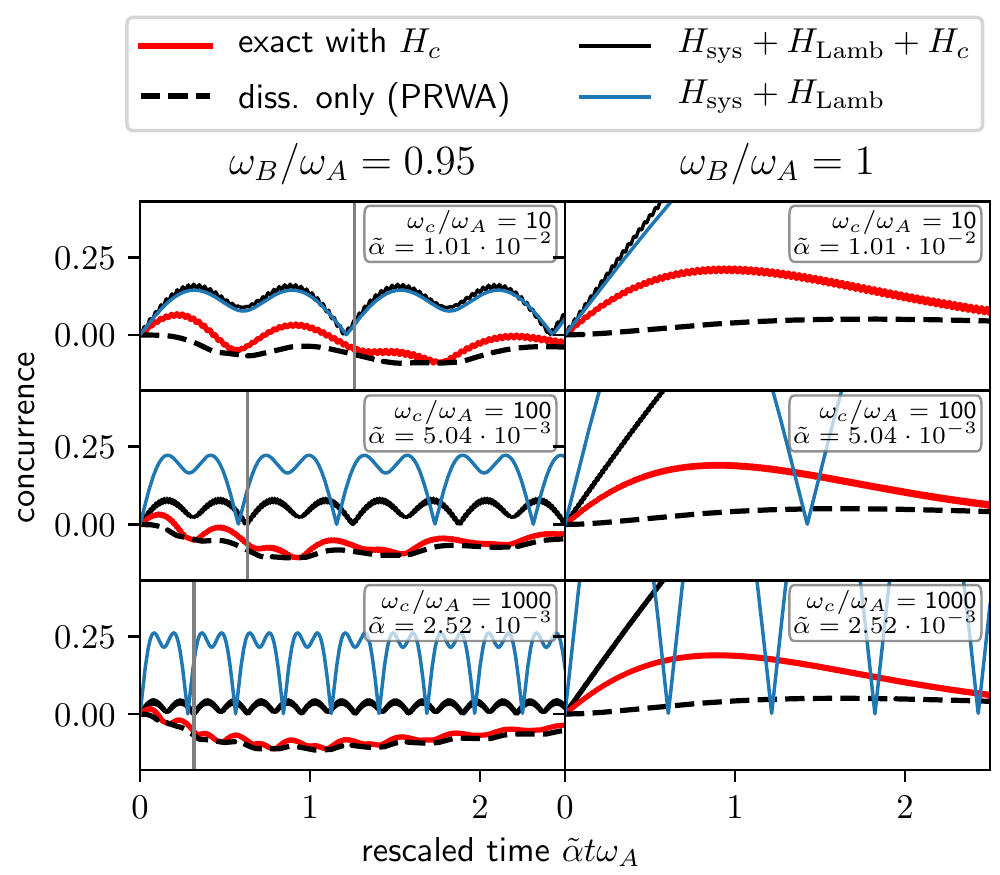}
  \caption{The entanglement dynamics (including the counterterm) is shown (red lines) for a sub-Ohmic environment with $s=0.3$ and $|S(0)|/\omega_A = 0.03 = \tilde \alpha \Gamma(s) \omega_c^s / \omega_A^s / 2$ while increasing the cutoff frequency $\omega_c$ from top to bottom, hence $\tilde \alpha$ decreases.
  As expected, for the detuned case increasing $\omega_c$ results in a more effective compensation of the Lamb shift contribution by the counterterm (black solid line).
  This explains that the exact entanglement dynamics (red line) approaches the dynamics obtained by the \ac{PRWA} master equation under the action of the dissipator only (dashed line).
  Remarkably, for resonant qubits (right column) this expectation is not fulfilled.
  In that case the joint unitary dynamics $H_\sys + H_\mathrm{Lamb} + H_c$ (black solid line) builds up entanglement on the same time scale as the dissipation takes place, independently of the cutoff frequency $\omega_c$.
  Consequently, the exact dynamics does not approach the dynamics of the \ac{QOME} under the action of the dissipator only (dashed line).
  This qualitative difference originates from the time scale set by the detuning which is independent of $\omega_c$ (gray vertical lines). Note, this time scale decreases on the rescaled time $\tilde \alpha t \omega_A$ while increasing $\omega_c$.
  Since the entanglement dynamics of the joint unitary part is periodic on that time scale, but the buildup takes place on the time scale set by $\Delta S \propto S_0$ (constant on the rescaled time), the generation of entanglement is effectively suppressed.}
  \label{fig:large_wc_small_s_1_0_3}
\end{figure}

Nonetheless, the above reasoning suggests that in the limit $\omega_c \rightarrow \infty$, where the approximation $S(\omega_i) \approx S(0)$ becomes exact, the counterterm should truly cancel the Lamb shift contribution also in the deep sub-Ohmic regime.
For two detuned qubits this behavior is confirmed by the example shown in the left column of Fig. \ref{fig:large_wc_small_s_1_0_3}.
The entanglement dynamics (red line) approaches the purely dissipative dynamics (dashed black line) when increasing the cutoff frequency.
The unitary effect of the Lamb shift $H_\mathrm{lamb}$ (blue line) is compensated by the counterterm $H_c$ (solid black line) where the degree of compensation increases with the cutoff frequency $\omega_c$.
Hence, it seems plausible that the entanglement induced by the Lamb shift Hamiltonian vanishes in the adiabatic limit $\omega_c \rightarrow \infty$.

Remarkably, this does not hold true for resonant qubits (see right column in Fig. \ref{fig:large_wc_small_s_1_0_3}).
To affirm that this effect remains even in the adiabatic limit we argue as follows.
The remaining Hermitian contribution $H_\mathrm{Lamb} + H_c$ scales in lowest order like
\begin{equation}
  \Delta S = S_0 - S(\omega_A) \sim S_0 \left(\frac{\omega_A}{\omega_c}\right)^s
\end{equation}
which sets the timescale on which the unitary interaction between the qubits induced by the full Hamiltonian [Eq. \eqref{eqn:hamiltonian_with_CT}] takes place.
Obviously, increasing $\omega_c$ while keeping $S_0$ constant increases that timescale which could mean an effective cancellation if the damping takes place on a faster time scale.
It turns out, however, that this is not the case since the damping rate $\gamma$, determined by the SD at the resonance frequency, scales with $\omega_c$ in the same manner.
This can be seen by expressing the coupling strength in terms of $S_0$ and expanding the SD in lowest order in $\omega_A/\omega_c$
\begin{equation}
  \gamma \sim J(\omega_A) \sim S_0 \left(\frac{\omega_A}{\omega_c}\right)^s \quad.
\end{equation}
In summary we find the counter-intuitive result that for two resonant qubits and a sub-Ohmic environment the effect of the Lamb shift contribution is not compensated by the counterterm, even in the limit $\omega_c \rightarrow \infty$ which is usually considered the adiabatic limit.
Consequently, entanglement generation beyond the purely dissipative contribution remains for any $\omega_c$.
This, in particular, stays in contrast with the results presented in Ref. \cite{KastBipartiteentanglementdynamics2014}.
The authors claim that for sufficiently weak coupling the model including the counterterm (as in Eq. \ref{eqn:hamiltonian_with_CT}) does not result in any entanglement generation for the two qubits.

\section{Conclusions}

In the context of an open two-qubit system we address the fundamental question: To what extent does the incorporation of the counterterm into the microscopic model counterbalance the effect of the Lamb shift Hamiltonian and, thus, inhibit the environmentally induced generation of entanglement?
The answer requires some prerequisites which we provided and discussed as well.
First of all we used the well tested \ac{HOPS} method to obtain the exact entanglement dynamics.
With the exact dynamics at hand we were able to assess the accuracy of various perturbative approaches.
We found that the \ac{RFE} yields the most accurate results which is not too surprising since it involves the least approximations. 
We argued that positivity violations can be consistently dealt with by using the positive matrix $\sqrt{\rho \rho^\dagger}$ to estimate the amount of entanglement.
However, to answer our main question a master equation of \ac{GKSL} kind is desirable which clearly separates the unitary from the dissipative influence of the environment.
As expected, we confirm that the \ac{GAME} as well as \ac{PRWA} approach yield reasonable entanglement dynamics also for detuned qubits.
Concerning the above assertion, if the Lamb shift Hamiltonian (unitary influence of the environment) is truly compensated by the counterterm the resulting dynamics should match the perturbative dynamics obtained by the master equation under the exclusive action of the dissipator in the weak coupling regime.
Therefore, we compared the exact dynamics including the counter term with the dissipator only dynamics of the \ac{PRWA} approach.
The \ac{GAME} approach could have been used equally well.
It is clear, that for finite $\omega_c$, if at all, the compensation is only approximate.
We found that for an Ohmic environment the exact dynamics matches fairly well the dissipative dynamics of the perturbative approach -- the counterterm compensates the Lamb shift Hamiltonian in some sense.
However, decreasing $s$ (sub-Ohmic environment) while keeping $\omega_c$ constant worsens the compensation. 
For small $s \lesssim 0.3$ the entanglement generation is rarely influenced by the counterterm.
This behavior can be traced back to the power law behavior of $S(\omega) \sim |\omega/\omega_c|^s$, the imaginary part of the half sided Fourier transform of the \ac{BCF}, and is amplified by the pointed minimum of $S(\omega)$ at $\omega=0$ for $s < 0.5$.
Remarkably, for resonant qubits and a sub-Ohmic environment even in the adiabatic limit $\omega_c \rightarrow \infty$ the entanglement generation cannot be modeled solely by the dissipator of the perturbative master equation.
We showed that on the one hand, increasing $\omega_c$ slows down the generation of entanglement, which is due to the still imperfect compensation.
On the other hand the relaxation time increases in precisely the same manner.
Consequently, even in the limit $\omega_c \rightarrow \infty$ the entanglement dynamics is influenced by the infinitesimal non-zero difference between the Lamb shift Hamiltonian and the counterterm.

In summary, for an environment with finite $\omega_c$ including the counterterm in the microscopic model does not imply that the unitary influence of the environment on the system is canceled.
For the general detuned case, in order to achieve the same level of compensation while going from the Ohmic to the deep sub-Ohmic regime the cutoff frequency $\omega_c$ needs to be increased.
It is a special property of two resonant qubits that even though the Lamb shift Hamiltonian and the counterterm approach each other in the large cutoff regime the entanglement dynamics is influenced significantly by the remaining infinitesimal difference.

Alongside these main findings we reported that in a perturbative regime the \ac{RFE} with time dependent coefficients yields the most accurate entanglement dynamics when curing positivity issues by the replacement $\rho \rightarrow \sqrt{\rho \rho^\dagger}$.
The recently proposed \ac{GAME} which guarantees positivity yields results which qualitatively agree with the \ac{RFE} while being slightly less accurate.
Applying the \ac{RWA} only partially yields reasonable entanglement behavior also for detuned qubits while superimposed fast oscillations are obviously missing.
As of the rather small cutoff frequency considered here and the slow algebraic decay of the \ac{BCF} the \ac{CGME} is not suitable.
These findings are in line with results in Ref. \cite{HartmannAccuracyAssessmentPerturbative2020} obtained for a Lorentzian environment.

In addition we report that although the initial entanglement generation is followed by a decay, the asymptotic value of entanglement does not vanish for the initial states $\ket{\uparrow \uparrow}$ and $\ket{\downarrow \downarrow}$.
From the exact numeric solution we find a linear relation between the asymptotic entanglement and the coupling strength.
None of the perturbative methods considered here capture this feature correctly.

Further, independent of the particular model and based on the \ac{RFE} we argued that the Lamb shift Hamiltonian and the counterterm cancel up to some error which is related to the approximation $S(\omega_i) \approx S(0)$ for all $\omega_i$ where $\omega_i$ takes the values of all possible transition frequencies of the system Hamiltonian.
For a (sub-) Ohmic environment this approximation becomes exact in the limit $\omega_c \rightarrow \infty$ (adiabatic limit).
This, however, does not necessarily imply that the dynamics is not influenced by the remaining very small unitary influence of the environment, as explained above.

As a continuation of the results presented here, the entanglement dynamics under the influence of a thermal environment, again with particular focus on the perturbative regime, will be considered in future work.
Both, the exact \ac{HOPS} method as well as the perturbative master equations are capable of treating non-zero temperature environments.
In general, as of the additional thermal fluctuations we expect less entanglement generation.
The particular behavior of the asymptotic entanglement and whether or not the anomaly in the adiabatic regime will remain has yet to be investigated.

We gratefully acknowledge fruitful discussions with Kimmo Luoma, Valentin Link, Konstantin Beyer and Dario Egloff.
We also thank the \mbox{IMPRS} "Many-Particle Systems in Structured Environments" for their support.
The computations were performed on a Bull Cluster provided at the Center for Information Services and High Performance Computing (ZIH) at TU Dresden.

\bibliographystyle{unsrtnat}
\bibliography{inducedEntanglement}

\end{document}